\documentclass[
    ,final 
  ,numberedheadings 
  ,cmfonts          
  ,float
  ,subfigure
  ]
  {aipproc}

\usepackage{amsmath}
\usepackage{subfigure}
\layoutstyle{6x9}

\newcommand{\cf}[1]{{Fig.~\ref{#1}}}

\newcommand{\eqs}[1]{\begin{equation} \begin{split} #1\end{split} \end{equation} }

\newcommand{\gmu}{\gamma^\mu}

\renewcommand{\thefootnote}{\fnsymbol{footnote}}

\begin{document}
\title{Heavy Quarkonium Production: Extending CSM and COM}

\author{J.P.~Lansberg}{
address={Physique th\'eorique fondamentale, D\'epartement de  Physique, Universit\'e de  Li\`ege, \\ 
all\'ee du 6 Ao\^{u}t 17, b\^{a}t. B5, B-4000 Li\`ege~1, Belgium\\
E-mail: JPH.Lansberg@ulg.ac.be}
}

\begin{abstract}
By questioning the applicability of the static approximation of the Colour-Singlet Model,
we have seen that  the production amplitude receives contributions from two different cuts. 
The first one in its static limit gives the colour-singlet mechanism. 
The second one has not been considered so far. We treat it in a gauge-invariant manner 
by introducing necessary new 4-point vertices, suggestive of the colour-octet
mechanism. This new contribution can be as large as the colour-singlet mechanism
at high $p_T$, however these vertices are not totally constrained and
when the freedom in their determination is fully exploited, we are able to reproduce the production
cross-sections at the Tevatron for the $J/\psi$, $\psi'$ and $\Upsilon(1S)$ and at RHIC for 
the $J/\psi$.
\end{abstract}

\maketitle


\footnotetext{Presented at DIS 2005, Madison, Wisconsin, 27 Apr - 1 May 2005 and at PHENO 05, 2-4 May 2005, Madison, Wisconsin.}
\renewcommand{\thefootnote}{\arabic{footnote}}

\section{Introduction}

Ten years after the discovery of the ``$\psi'$ anomaly'' by the CDF 
  collaboration~\cite{CDF7997a,CDF7997b}, 
no totally conclusive solution has been proposed so 
far (for a comprehensive and up-to-date review on the subject, see~\cite{Brambilla:2004wf}). Even though the Colour-Octet 
Mechanism (COM), coming from the application of NRQCD to heavy quarkonium, is a good 
candidate, it appears clearly that as long as fragmentation is the dominant
production contribution and the velocity scaling rules of NRQCD hold, it cannot accommodate 
the polarisation measurements of CDF~\cite{Affolder:2000nn}, which show a non-polarised, if not 
slightly longitudinal, production.

In that context, we have felt the necessity to reconsider the appropriateness of 
the static and on-shell approximation of the Colour-Singlet Model (CSM)~\cite{CSM_hadron}, 
which is still the most natural model from QCD. These approximations are also
implicit in the COM, therefore any feature arising from this study should 
have some implication for the COM. 

In order to study properly non-static and off-shell effects, we have used a vertex function
as an input for the bound state characteristics, whereas the Schr\"odinger wave function at the origin 
is used in the CSM and Long Distance Matrix Elements (LDME) of NRQCD enter the COM. We emphasise again
that we probe all the internal phase space of the quarkonium, and thus need a function, where the two
models simply need a constant factor.

\section{Our Model}

In the case of $^3S_1$ quarkonium (noted $\cal Q$) production in 
high-energy hadronic collisions, we are to consider gluon fusion $gg\to {\cal Q} g$. 
Using the Landau equations~\cite{Landau:1959fi}, we have shown in~\cite{Lansberg:2005pc} that 
there are two families of contributions (see~\cf{fig:diag_LO_QCD} (a) and (b)): 
the first is the usual colour-singlet mechanism, where in the context of our model, 
we use a 3-point function $\Gamma^{(3)}_{\mu}(p,P) = \Gamma(p,P) \gamma_\mu$
 at the $Q\bar Q \cal Q$ vertex; the second family was never considered before. 
To simplify the study, we set $m>M/2$ so that the first cut does not contribute.

For the functional form of $\Gamma(p,P)$, we neglect
possible cuts, and choose two opposite scenarios:
a dipolar form which decreases gently with its argument, and a Gaussian form:
$\Gamma(p,P)=N(1+\frac{\vec p^{\, 2}}{\Lambda^2})^{-2}$ and 
$\Gamma(p,P)=N e^{-\frac{\vec p^{\, 2}}{\Lambda^2}}$, both with a 
free size parameter $\Lambda$, and a normalisation $N$. In~\cite{proc_spa}, we 
have shown how to fix the normalisation $N$ of $\Gamma(p,P)$ as a function of $\Lambda$.

\begin{figure*}
\centerline{\mbox{\includegraphics[height=4cm]{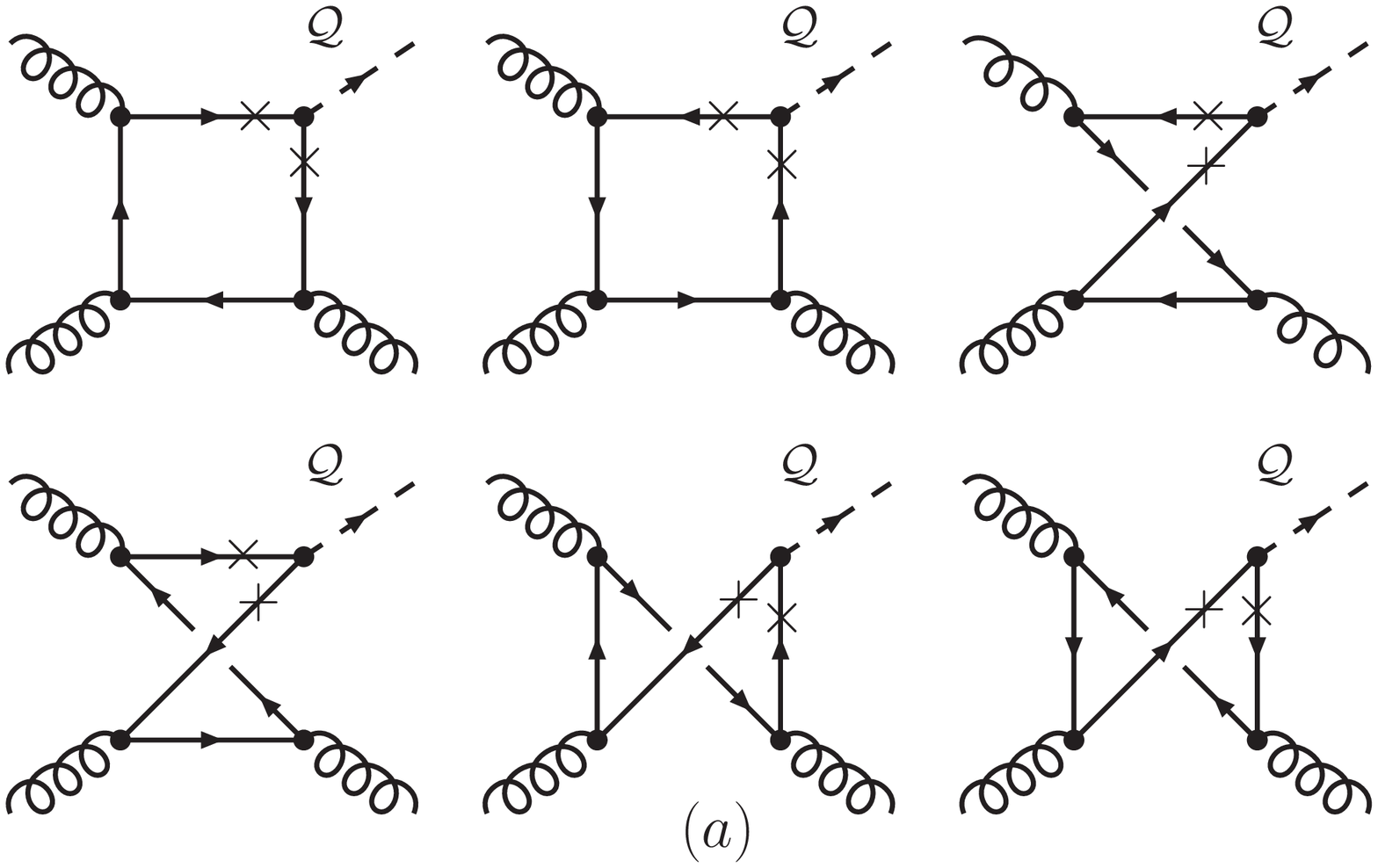}\quad
\includegraphics[height=4cm]{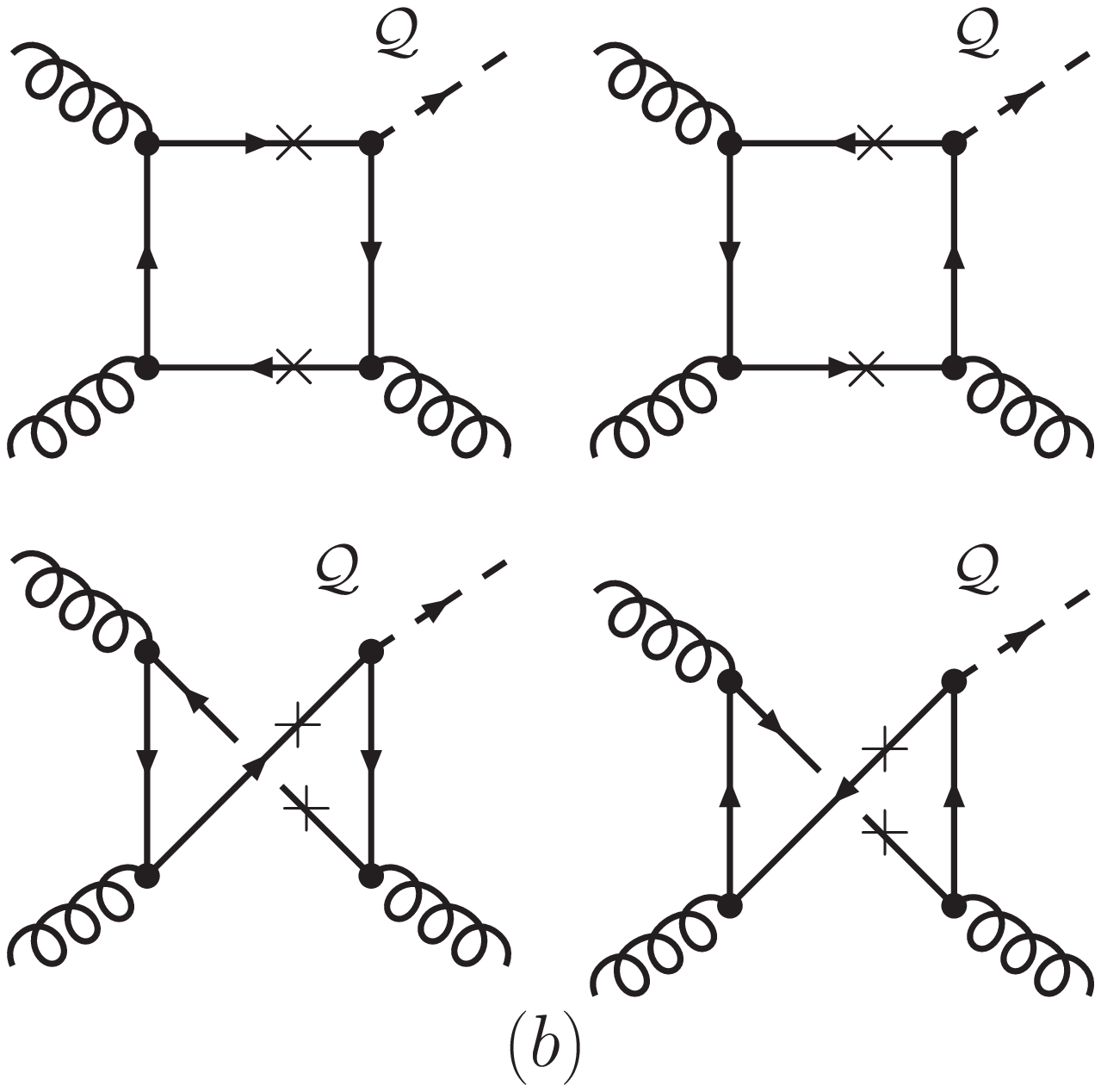}
\quad\includegraphics[width=3cm]{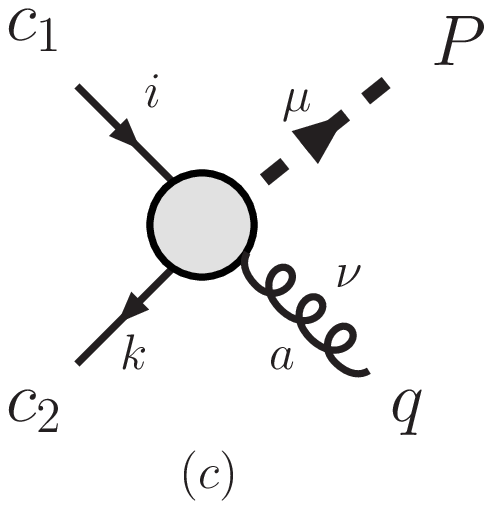}}}
\caption{The first family (a) has 6 diagrams and  the second family (b)  
4 diagrams contributing the discontinuity of $gg \to \!\!\ ^3S_1 g$ at LO in QCD.(c): the gauge-invariance restoring vertex, $\Gamma^{(4)}$.}
\label{fig:diag_LO_QCD}
\end{figure*}

In addition to the second family, one is driven -- to preserve gauge invariance (GI) -- to introduce
new contributions arising from the presence of 4-point vertices. Besides restoring
GI, these vertices have to satisfy specific constraints~\cite{Lansberg:2005pc,these,Drell:1971vx}. 
For the following 
simple choice for $\Gamma^{(4)}_{\mu\nu}(c_1,c_2,P,q)$
\begin{eqnarray}
-i g_s T^a_{ki} \left( \Gamma(2c_1-P,P)
-\Gamma(2c_2-P,P)\right)\left[{c_{1\mu}\over (c_2+P)^2-m^2}
+{c_{2\mu}\over (c_1-P)^2-m^2}\right]\gamma_\nu,
\end{eqnarray}
where the momenta and indices are as in~\cf{fig:diag_LO_QCD} (c), we got for the $J/\psi$ and $\psi'$ 
production at the Tevatron the results shown in~\cf{fig:cross-section_01}. In the $\psi'$ case, 
we employed the ambiguity upon the vertex function normalisation due to the node position $a_{node}$ to 
describe the data at low $P_T$. Note that the slope is not that different from that of the data. 
This is at variance with what is widely believed since fragmentation 
(with a typical $1/P_T^4$ behaviour) processes describe the data.

\begin{figure*}
\centering
\includegraphics[height=4.75cm,clip=true]{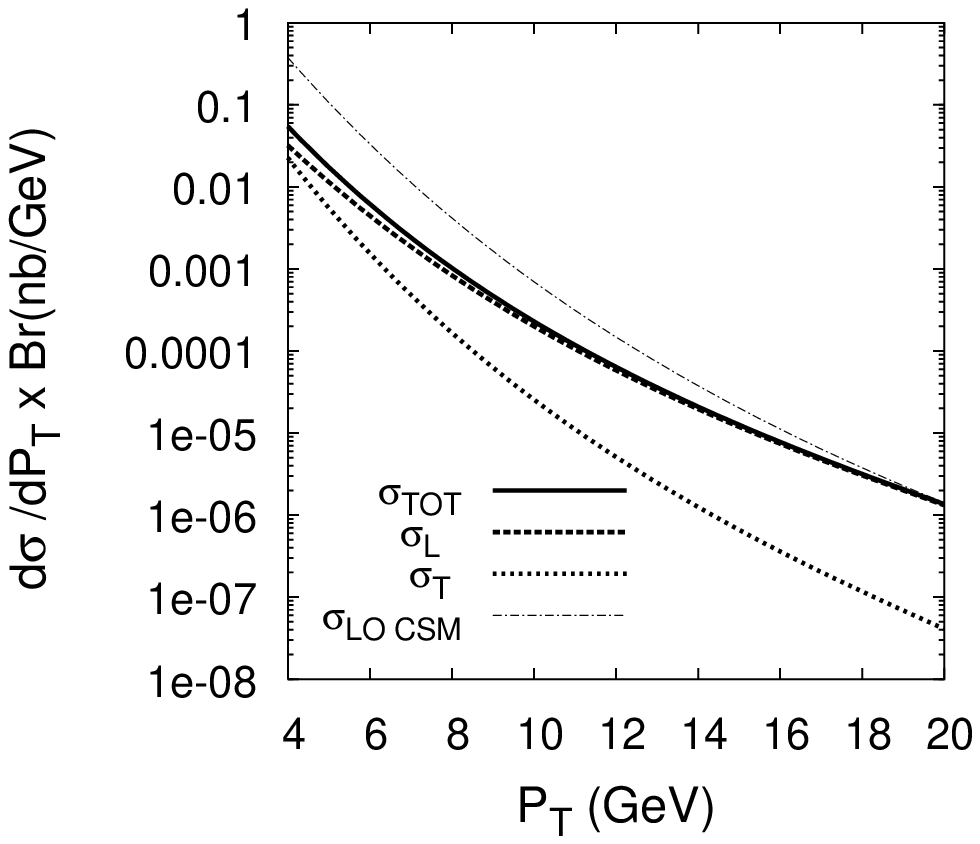}
\includegraphics[height=4.5cm,clip=true]{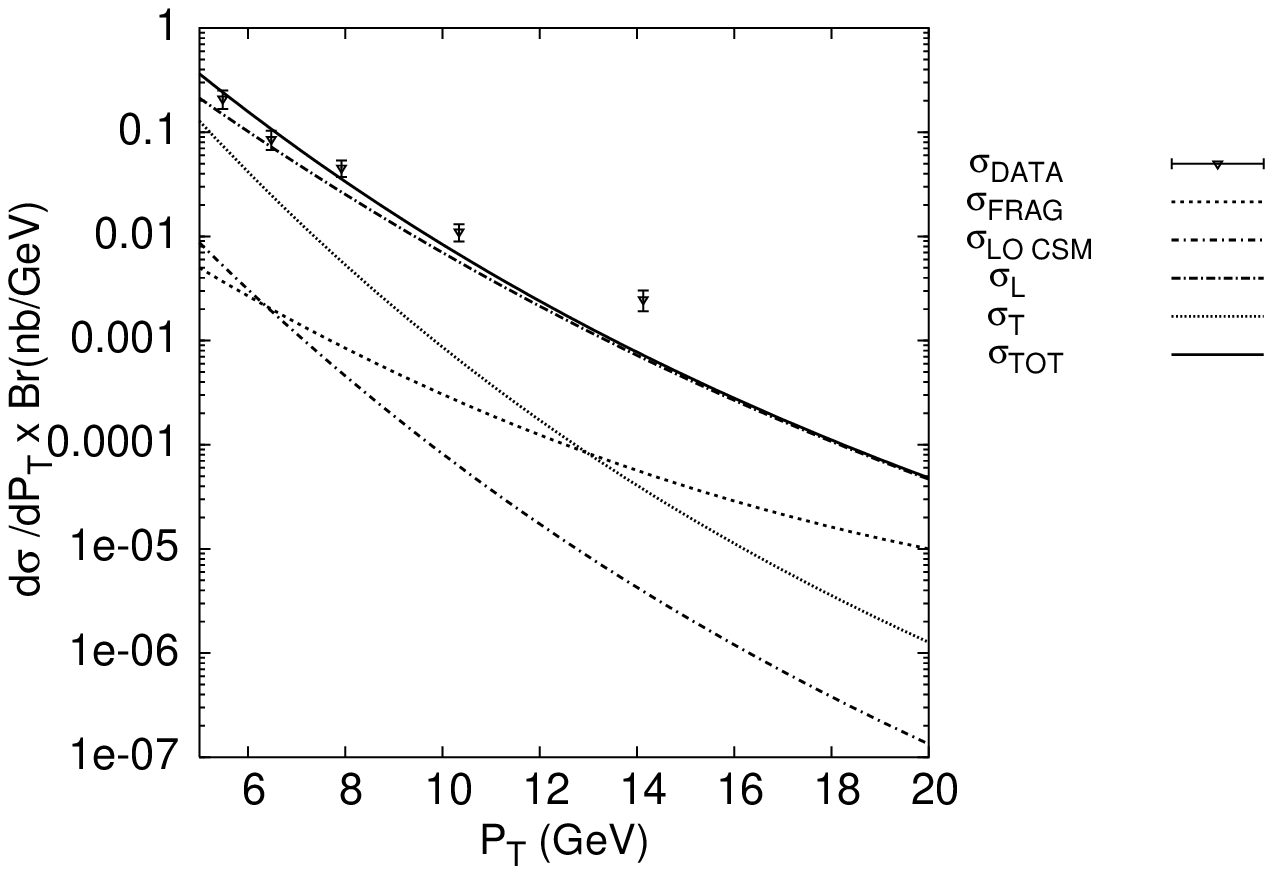}
\caption{Polarised ($\sigma_T$ and $\sigma_L$) and total ($\sigma_{TOT}$) cross sections obtained 
with a Gaussian vertex functions On the left, for $J/\psi$ with $m=1.87$ GeV and $\Lambda=1.8$ GeV; 
on the right for $\psi'$ with $a_{node}=1.333$ GeV, $m=1.87$ GeV and $\Lambda=1.8$ GeV 
to be compared with the data from CDF~\cite{CDF7997a,CDF7997b}.
}
\label{fig:cross-section_01}
\end{figure*}

However there exist different choices for the GI restoring vertex (GIRV). In the following, we present 
some interesting results obtained by studying the effects of autonomous vertices. The latter link
different suitable choices of GIRV: they are GI alone and a priori unconstrained in normalisation. 
For a first study, let us restrict the choice to the three simplest possible 
ones~\cite{these,article2} (omitting the factor $-i g_s T^a_{ki} 
\left( \Gamma(2c_1-P,P)-\Gamma(2c_2-P,P)\right)$)
\eqs{
(a) \ \alpha/(\sqrt{\hat s}m_{\cal Q}) \gmu q^\nu \hspace*{1.5cm} (b) \ \beta' \, 
(c_1+c_2)^\mu (c_1+c_2)^\nu \hspace*{1.5cm} (c) \ \xi/m_{\cal Q} \, g^{\mu\nu}
}

The factors $\alpha$, $\beta'$ and $\xi$ are  free constant. If we 
introduce these contributions in the amplitude calculation, we see in~\cf{fig:g_m528_l600_cteq_10-8}
 that we can fit the data for some set of values for ($\alpha$, $\xi$).

\begin{figure*}
\parbox{15cm}{\includegraphics[height=5cm]{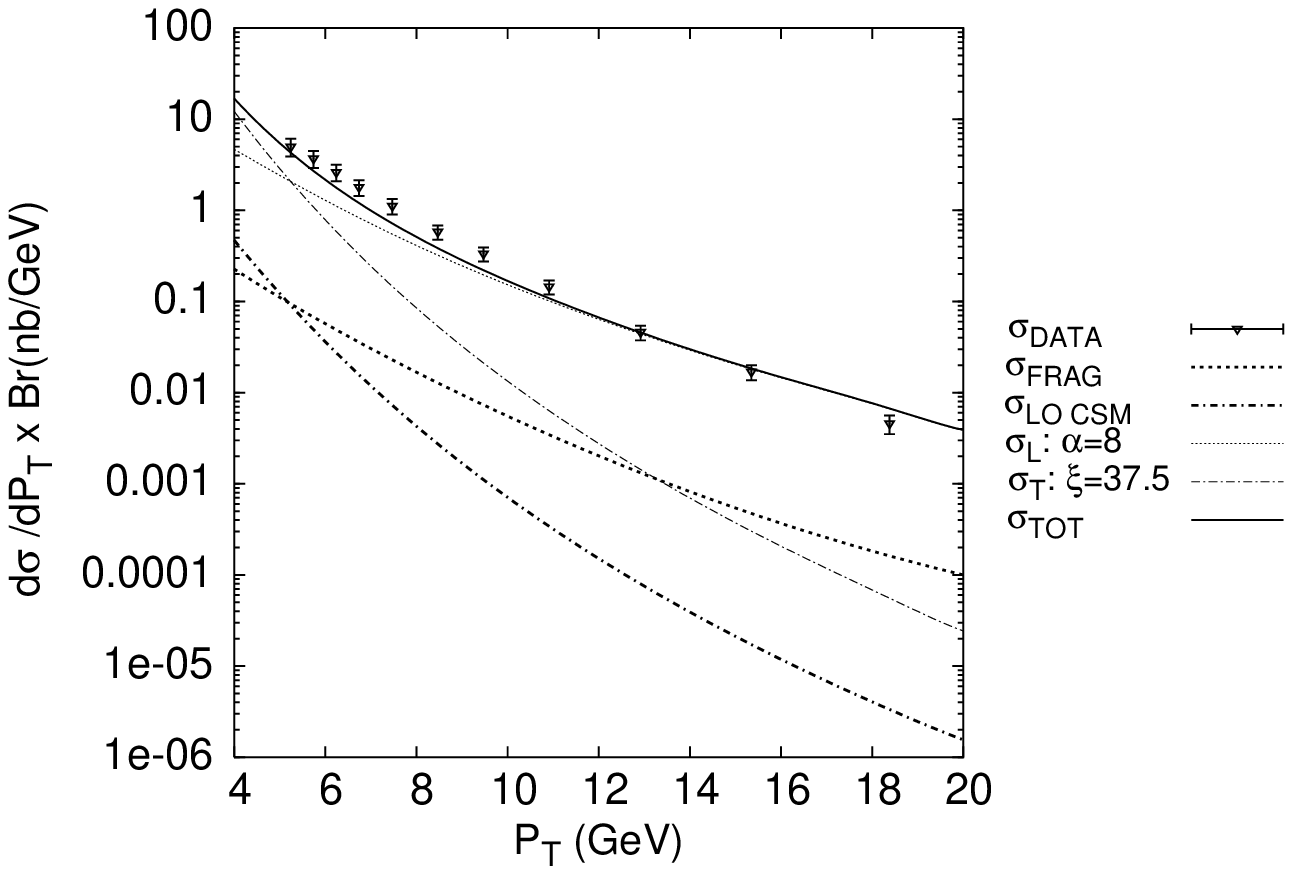}
\includegraphics[height=5cm]{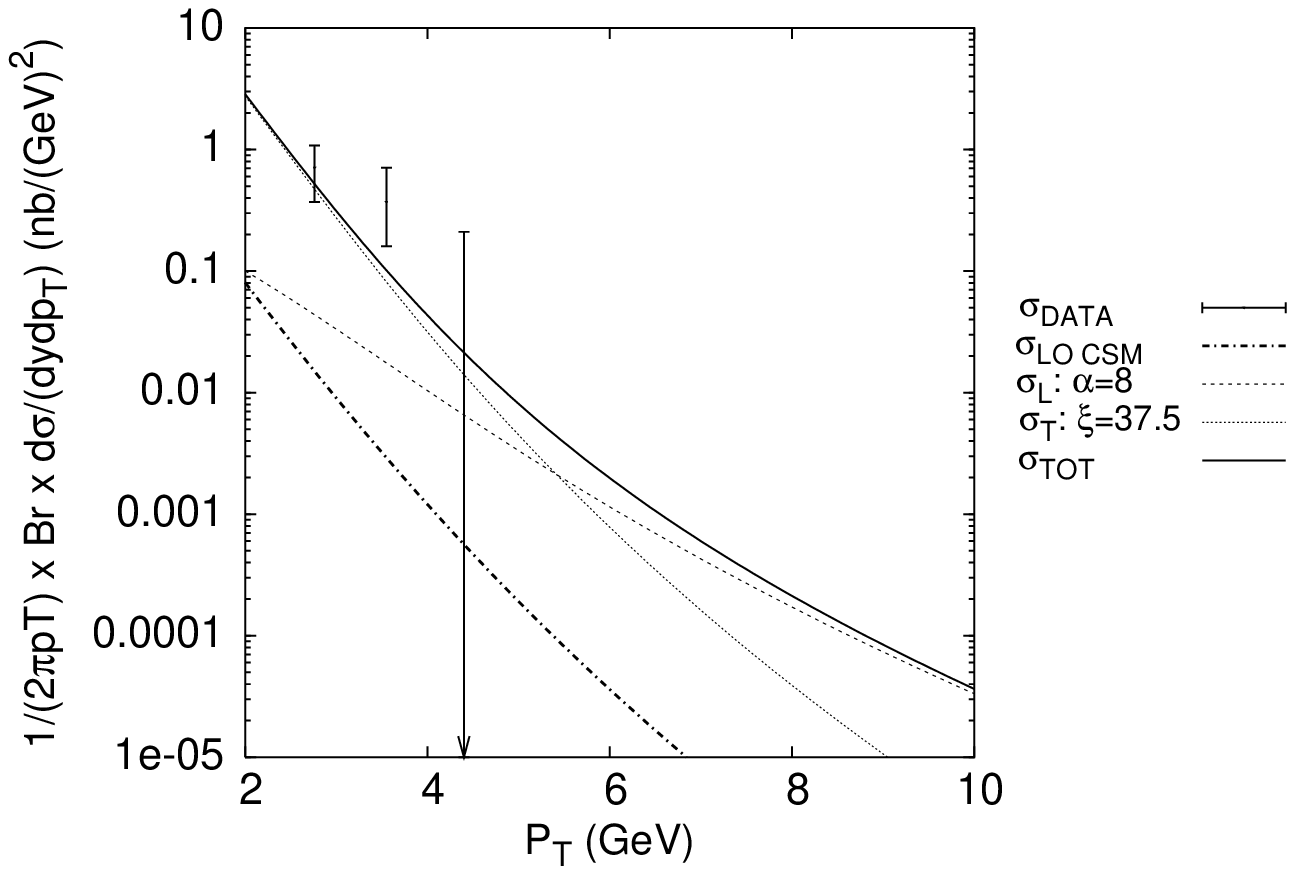}
\includegraphics[height=5cm]{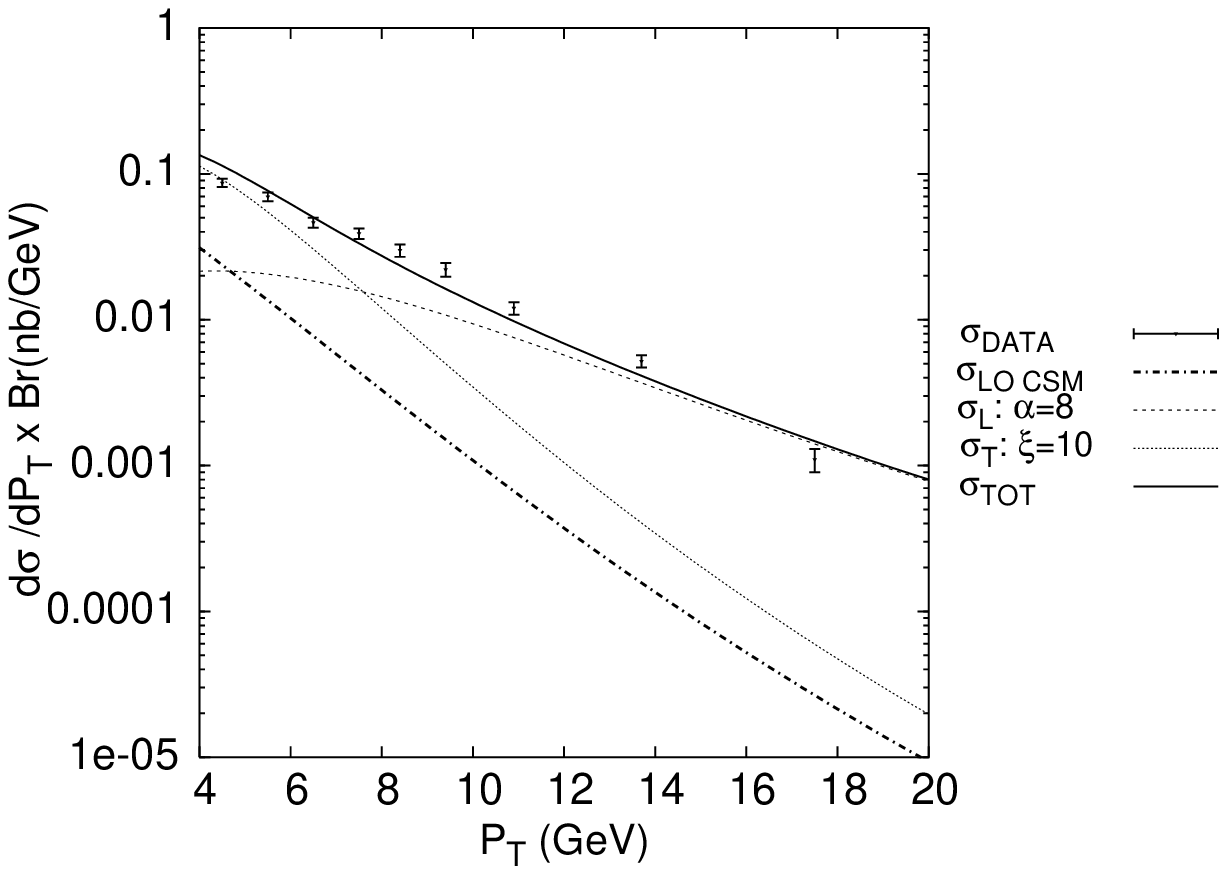}
\includegraphics[height=5cm]{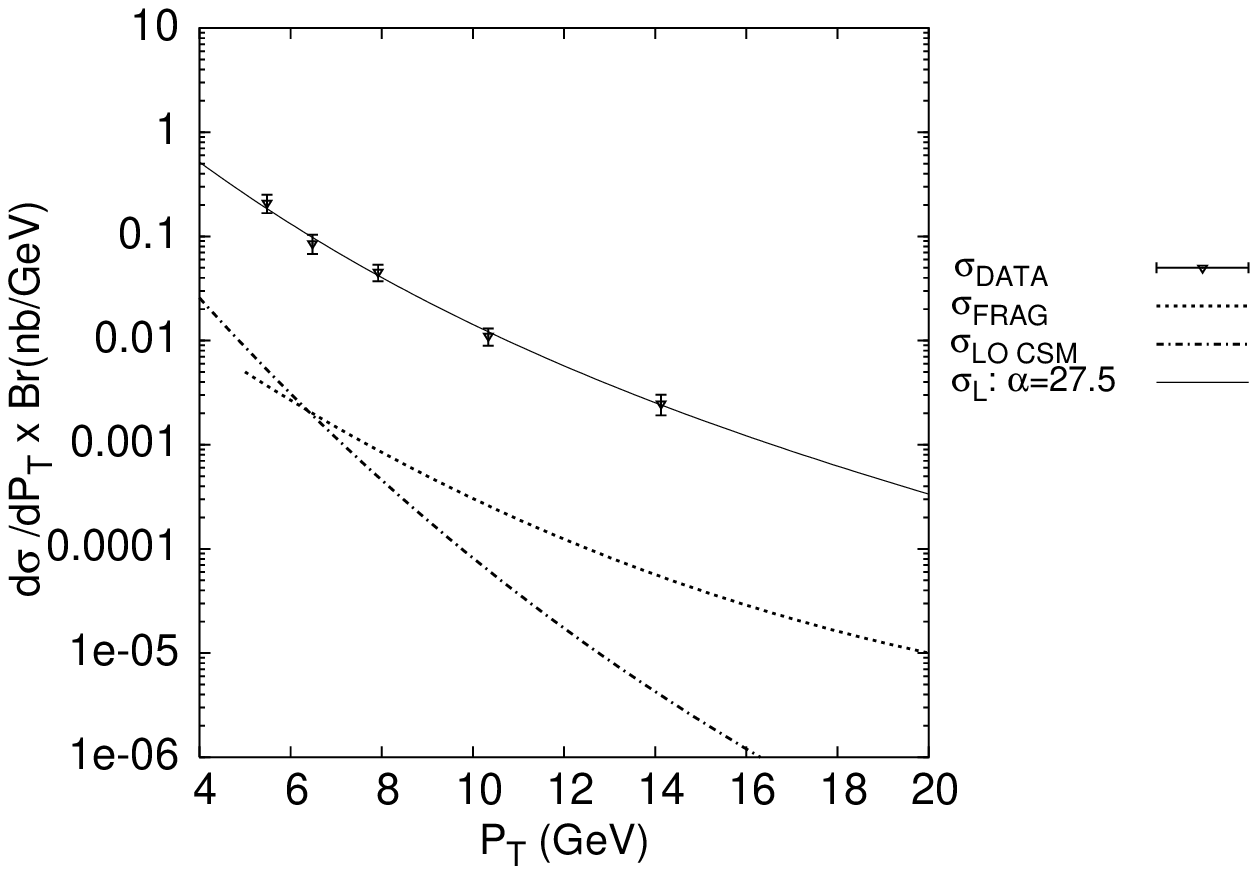}}
\caption{Polarised ($\sigma_T$ and $\sigma_L$) and total ($\sigma_{TOT}$) cross sections 
obtained: upleft -- for $J/\psi$ at $\sqrt{s}=1800$~GeV with $\alpha=8$ and $\xi=37.5$ 
to be compared with LO CSM, the fragmentation 
in the CSM and with the data of CDF~\cite{CDF7997b}; upright -- 
for $J/\psi$ at $\sqrt{s}=200$~GeV with $\alpha=8$ and $\xi=37.5$ to be compared  with LO CSM and 
with the data of PHENIX~\cite{Adler:2003qs}; downleft -- for $\Upsilon(1S)$  at 
$\sqrt{s}=1800$~GeV with $\alpha=8$ and $\xi=10$ to be compared with LO CSM and with 
the data of CDF~\cite{Acosta:2001gv}; downright -- for $\psi'$ at 
$\sqrt{s}=1800$~GeV with $\alpha=27.5$ and to be  compared with LO CSM, 
the fragmentation in the CSM and with the data of CDF~\cite{CDF7997a}.}
\label{fig:g_m528_l600_cteq_10-8}
\end{figure*}

\section{Conclusion}

We have shown that it is possible to go beyond the static approximation of the CSM. It 
may also be possible to extend the COM in the same manner. This necessitates the 
introduction of 4-point vertices due to the non-local 3-point vertex relevant for the non-static and
off-shell contributions.

By going deeper in the analysis, we see that the form of these 4-point vertices is not absolutely 
constrained even after imposing necessary conditions to conserve crossing symmetry and the analytic 
structure of the amplitude. When this lack of constraint is used, we
are able to reproduce the cross-section for the $J/\psi$, $\psi'$ and $\Upsilon(1S)$ 
as measured at the Tevatron by CDF (and also at RHIC by PHENIX for $J/\psi$).

In our framework, cross-sections are dominated by longitudinal $\cal Q$, 
therefore by combining our approach with COM fragmentation they could~\cite{article2} agree with 
the polarisation measurements of CDF.


\begin{theacknowledgments}
J.P.L. is an IISN Postdoctoral Researcher, this work
 has been done in collaboration with J.R.~Cudell and Yu.L. Kalinovsky~\cite{Lansberg:2005pc}. 
\end{theacknowledgments}

\bibliographystyle{aipprocl} 

\end{document}